\documentclass[english,keywords,showpacs,amsmath,amssymb,preprint]{revtex4}
\usepackage[T1]{fontenc}
\usepackage[latin1]{inputenc}
\usepackage{babel}%
\usepackage{graphicx}
\usepackage{color}
\usepackage{bm}
\usepackage{longtable}
\usepackage{amsmath}
\usepackage{amsfonts}
\usepackage{amssymb}

\begin{document}
\title{Weak and semiweak values in non-ideal spin measurements: an exact treatment beyond the asymptotic regime}
\author{A. K. Pan and A. Matzkin}
\affiliation{Laboratoire de Physique Th\'{e}orique et Mod\'{e}lisation (CNRS Unit\'{e} 8089), Universit\'{e} de Cergy-Pontoise, 95302
Cergy-Pontoise cedex, France}

\begin{abstract}
We consider weak measurements (WM) of a spin observable in quantum mechanics beyond the usual asymptotic regime. This is done by obtaining the exact time-dependent wave functions of the measuring apparatus for general non-ideal measurements. Ideal strong measurements and the usual WM regime are obtained as the two extreme limiting cases of our exact treatment. We show that in the intermediate regime non-ideal measurements lead to ``semiweak'' eccentric values that differ from the usual weak values. We further show that even in the WM regime the exact treatment leads to a meter behavior that can be markedly distinct from the one predicted by the usual WM formalism. We give several illustrations and discuss an application to the distinction of different realizations of the same density matrix.

\end{abstract}


\maketitle

\section{Introduction}

In recent times the issue of `weak measurement'(WM) in quantum mechanics (QM)
has gained a significant and wide interest in realizing apparently
counterintuitive quantum effects. This path-breaking idea was originally
proposed by Aharonov, Albert and Vaidman(AAV) \cite{aav} that introduced a
measurement scenario in QM so that the empirically measured value (coined as
`weak value') of an observable can be seemingly weird in that it yields
results going beyond the eigenvalue spectrum of the measured observable. Since
then, this idea has been enriched by a number of theoretical works
\cite{duck,av91,mit,jeff,jozsa,brunner} and
experimental works
\cite{ritchie,wang,pryde,hosten,yokota,starling,lundeen09,lund,zil}
This novel technique of obtaining strange property of an observable has
several implications, for example, to provide insights into conceptual quantum
paradoxes \cite{av91,vaidman,aha02,mit}, identifying a tiny spin Hall
effect\cite{hosten}, detecting very small transverse beam deflections
\cite{starling} and measuring average
quantum trajectories for photons \cite{steinberg11}.

As it is well known a standard quantum mechanical measurement requires a
\emph{strong} coupling in order to produce an \emph{ideal} one-to-one
correspondence between the measured system and the device state (for this
reason standard measurements are often coined as being \emph{strong} or
\emph{ideal}). In contrast the WM scheme, as the name suggests, assumes a very
weak coupling (so that the system state is kept grossly undisturbed) and
interfering device states (so that the coherence in the device remains
intact). To observe an effect of this weak perturbation in the device
pointer's state, it is necessary to suitably select a particular subensemble
which is technically termed as ``post-selection''. Given the initial state
(known as ``pre-selected'' state in AAV's terminology) any `strange' weak
value of an observable can be observed if the post-selection is appropriately
chosen. The weak values have another unusual property in that they can be
complex. If the weak coupling approximation is made in position coordinate,
such as, in a Stern-Gerlach (SG) setup, the real and imaginary parts of the
complex weak value correspond to the momentum and position mean values of the
post-selected pointer state respectively \cite{jozsa}. In this case, if the
weak value is real the effect can be observed in the momentum space
distribution of the pointer state, i.e., the peak of the momentum distribution
is shifted by an amount proportional to the weak value, possibly several times
the corresponding eigenvalue of the observable in question.

In the original AAV treatment, the weak measurement scenario is obtained as an
approximate limiting case for weak couplings and nearly overlapping device states, whereas
strong projective measurements pertain to the opposite limiting case. However,
between these two limiting cases there is a whole continuous range of
situations encompassing different regimes that have been scarcely
investigated.  Recently however, a few works \cite{geszti,wu,nakamura,zhu}
have studied the issue of weak measurement beyond the usual approximations
made in the AAV formalism. The specificity of our work is that we analyze
the general case of non-ideal measurements (followed by a post-selection) by starting from
the \emph{exact} solutions of the Schr\"{o}dinger equations coupling the system and the measuring device.
The standard WM regime then appears for weak couplings and nearly overlapping meter states.
But the other  ``non-ideal'' measurement situations also display potentially interesting effects.
We will show in particular that in these regimes
not only can eccentric outcomes be obtained, but that the
resulting ``semiweak'' or ``generalized weak''values can be tuned at will, allowing the device's pointer
to be shifted more than the weak value. We will also see that it is possible
to observe dichotomic outcomes of the pointer, as in the case of strong
measurements, but with eccentric shifts, including in situations in which
standard weak values cannot be defined. We will further show that even
in the standard WM regime, some effects neglected in the AAV treatment can
lead to exact weak values markedly different from the usual WM ones.
All this features will be analyzed in a simple
system that is analytically tractable - a spin-1/2 particle passing through a
series of Stern-Gerlach setups.

The paper is organized as follows. In Sec.\ II we briefly recapitulate the
essence of the standard weak measurement. In Sec.\ III we will introduce a
particle with spin 1/2 passing through a series of Stern-Gerlach setups. The
spatial wave packet will be considered as the probe that measures the
particle's spin state. Solving rigorously the coupled solutions of the system
allows to define naturally the non-ideal situation; the two limiting cases of
strong and weak measurements will be explicitly given; the peculiar features
visible in the intermediate regime -- the semi-weak and the generalized weak
values -- will be described. Our findings are illustrated in Sec.\ IV, and an
 application to a quantum information task will also be considered.
Our conclusions will be given in Sec.\ V.

\section{Standard weak measurement scenario}

The entire process of the AAV weak measurement procedure \cite{aav,av91} consists
of three different steps: state preparation (usually termed as pre-selection),
a strong projective measurement for selecting a specific subensemble known as
post-selection and in between the pre- and post-selection a weak interaction
is introduced so that system state remains virtually unaffected by this
intermediate interaction (see Fig.\ 1). AAV demonstrated the entire process in
terms of a series of three Stern-Gerlach (SG) setups. Let a beam of spin 1/2
neutral particles, say neutrons, pass through the SG setups. The first SG is
used to prepare the spin pre-selected state labeled by $|\chi_{in}\rangle$.
The total initial wave function after the first SG setup is $\Psi_{in}%
=\psi_{0}(x)|\chi_{in}\rangle$. The spatial part $\psi_{0}({x})$ is taken to
be a Gaussian wave packet peaked at the entry point (${x}=0$) of the second SG
at $t=0$
\begin{equation}
\psi_{0}(x)=\frac{1}{\left(  2\pi\delta^{2}\right)  ^{1/4}}\exp\left[
-\frac{x^{2}}{4\delta^{2}}\right]  .\label{inwfn}%
\end{equation}
For simplicity the spatial part is written as being one dimensional -- the
spatial wavefunction along the $y$ and $z$ directions are trivial in that they
are not affected by the Stern-Gerlach interaction given below. The initial
momentum space wave function $\phi_{0}(p_{x})$ corresponding to
Eq.(\ref{inwfn}) is%
\begin{equation}
\phi_{0}(p_{x})=\left(  \frac{2\delta^{2}}{\pi\hbar^{2}}\right)  ^{1/4}%
\exp\left(  -\frac{\delta^{2}p_{x}^{2}}{\hbar^{2}}\right)  .\label{a1}%
\end{equation}

The neutrons having the state $\Psi_{in}=\psi_{0}(x)|\chi_{in}\rangle$ then
pass through the second SG setup that is used for measuring a spin observable,
say, $\widehat{\sigma}_{x}$. The interaction Hamiltonian is given by
$H=f(t)\mu\widehat{\sigma}.\mathbf{B}$
where $\mathbf{B}=(bx,0,0)$ and $\mu$ is the magnetic moment of neutron.
$f(t)$ is a smooth function of $t$ vanishing outside the
interval $0<t<\tau$ and obeying $\int_{0}^{\tau}f(t)dt=\tau$,
where $\tau$ is the transit time during which the neutrons interact with the
magnetic field. The
total state after the interaction can then be written as%
\begin{equation}
\Psi^{\prime}=e^{-\frac{i\mu b\tau x{\hat \sigma}_{x}}{\hbar}}\psi_{0}(x)|\chi
_{in}\rangle.\label{a2}%
\end{equation}
The spatial part of the wavefunction can be considered in the context of the
SG setup as describing the state of the measurement device: in a SG, the spin
state is inferred from the wavepacket's deviation, that leads the particle to
be found int the upper or lower planes when a strong measurement is performed.

Here instead the magnetic field and transit time are taken to be very small,
so that the system state remains essentially unaltered by the interaction.
Then the exponential in Eq. (\ref{a2}) can be expanded to first order. This is
done in the original AAV treatment by performing, after the weak interaction,
a strong projective measurement by using the third SG setup. This
"post-selects" the neutrons in a definite final spin state $|\chi_{f}\rangle$.
The post selection allows to write the device state as%
\begin{equation}
\label{a5}
\psi_{f}(x)=\langle\chi_{f}|e^{-\frac{i\mu b\tau x{\sigma}_{x}}{\hbar}}%
\psi_{0}(x)|\chi_{in}\rangle=\langle\chi_{f}|1-i\frac{\mu b\tau x{\sigma}_{x}%
}{\hbar}+o(2)-...|\chi_{in}\rangle\psi_{0}(x)
\end{equation}
Neglecting the higher order terms leads to%
\begin{equation}
\psi_{f}(x)=\left\langle \chi_{f}|\chi_{in}\right\rangle \left[  1-\frac{i\mu
b\tau x}{\hbar}\frac{\left\langle \chi_{f}|{\sigma}_{x}|\chi_{in}\right\rangle
}{\left\langle \chi_{f}|\chi_{in}\right\rangle }\right]  \psi_{0}(x)\label{a6}%
\end{equation}
which can be written as
\begin{equation}
\psi_{f}(x)=\left\langle \chi_{f}|\chi_{in}\right\rangle e^{-i\frac{\mu b\tau
x}{\hbar}(\sigma_{x})_{w}}\psi_{0}(x)\label{a7}%
\end{equation}
where
\begin{equation}
(\sigma_{x})_{w}=\frac{\left\langle \chi_{f}|{\sigma}_{x}|\chi
_{in}\right\rangle }{\left\langle \chi_{f}\right\vert \chi_{in}\rangle
}\label{aavwvalue}%
\end{equation}
is known as weak value of the observable $\widehat{\sigma}_{x}$.

This weak value appears as a phase-shift in the pointer state's configuration
space wavefunction determines the position of the pointer in momentum space.
Taking the Fourier transform of Eq.(\ref{a7}) the pointer state in momentum
space can be written as%
\begin{equation}
\phi_{f}(p_{x})=\left\langle \chi_{f}|\chi_{in}\right\rangle \left(
\frac{2\delta^{2}}{\pi\hbar^{2}}\right)  ^{1/4}\exp\left[  -\frac{\delta
^{2}(p_{x}-p_{x}^{\prime}(\sigma_{x})_{w})^{2}}{\hbar^{2}}\right] \label{a10}%
\end{equation}
where $p_{x}^{\prime}=\mu b\tau$. The final momentum distribution can then be
written as
\begin{equation}
|\phi_{f}(p_{x})|^{2}=|\left\langle \chi_{f}|\chi_{in}\right\rangle |^{2}%
|\phi_{0}(p_{x}-p_{x}^{\prime}(\sigma_{x})_{w})|^{2}\label{a11}%
\end{equation}
where $|\left\langle \chi_{f}|\chi_{in}\right\rangle |^{2}$ is the probability
of successful post-selection. Hence the final pointer position in momentum
space is shifted by an amount $(\sigma_{x})_{w}p_{x}^{\prime}$ in contrast to
the strong measurement case where the final pointer positions are shifted by
$\pm p_{x}^{\prime}$, corresponding to the eigenvalues $\pm1 $ of $\hat
{\sigma}_{x}$.

From Eq.(\ref{aavwvalue}) it can be seen that $(\sigma_{x})_{w}$ can be
exceedingly large if pre-selected and post-selected state are nearly
orthogonal. Rescaling the momentum to $p_{x}/p_{x}^{\prime}$, we see the
pointer displays a broad distribution centered on $(\sigma_{x})_{w}$ (or the
real part of $(\sigma_{x})_{w}$ if the latter is complex), hence possibly well
beyond the ranges of the eigenvalues. For instance for a specific pre-selected
state, say, $|\chi_{in}\rangle=|\uparrow_{\theta}\rangle\equiv\cos\frac
{\theta}{2}|\uparrow_{z}\rangle+\sin\frac{\theta}{2}|\downarrow_{z}\rangle$
and post-selected state, say, $|\chi_{f}\rangle=\left\vert \uparrow
\right\rangle _{z}$, the weak value of $\hat{\sigma}_{x}$ is given by
$(\sigma_{x})_{w}=\tan\frac{\theta}{2}$. According to this formalism,
$(\sigma_{x})_{w}$ can become arbitrarily large as $\theta\rightarrow\pi$, but
simultaneously the probability of this post-selection, $|\left\langle \chi
_{f}|\chi_{in}\right\rangle |^{2}=|\langle\uparrow_{z}|\uparrow_{\theta
}\rangle|^{2}=\cos^{2}\frac{\theta}{2},$ becomes vanishingly small (we will
see below from the exact treatment that both of these properties are
erroneous, as the limit $\theta\rightarrow\pi$ is well-defined). The final
momentum distribution (\ref{a11}) after post-selection can be written as
\begin{equation}
|\phi_{f}(p_{x})|^{2}=\cos^{2}\frac{\theta}{2}\left(  \frac{2\delta^{2}}%
{\pi\hbar^{2}}\right)  ^{1/2}\exp\left[  -\frac{2\delta^{2}(p_{x}-p_{x}%
^{\prime}\tan\frac{\theta}{2})^{2}}{\hbar^{2}}\right]  .\label{aavpdis}%
\end{equation}

Note that in deriving the above relations we neglected the higher order terms
in Eq.(\ref{a5}) which is justified \cite{duck} if for $n\geq2$,%
\begin{equation}
\left(  \frac{\mu b\tau}{\hbar}\right)  ^{n}|x^{n}\left\langle \chi_{f}%
|(\hat{\sigma}_{x})^{n}|\chi_{in}\right\rangle |<<|\left\langle \chi_{f}%
|\chi_{in}\right\rangle |\nonumber
\end{equation}%
\begin{equation}
\left(  \frac{\mu b\tau}{\hbar}\right)  ^{n}|x^{n}\left\langle \chi_{f}%
|(\hat{\sigma}_{x})^{n}|\chi_{in}\right\rangle |<<\frac{\mu b\tau}{\hbar
}|x\left\langle \chi_{f}|\hat{\sigma}_{x}|\chi_{in}\right\rangle |\nonumber
\end{equation}
Moreover in order to resum Eq.(\ref{a6}) into Eq.(\ref{a7}) we need $\mu b\tau
x(\sigma_{x})_{w}/\hslash<<1;$ lastly noting that $x$ is effectively governed
by the spread $\delta$ we may write the condition for the derivation as%
\begin{equation}
\delta\frac{\mu b\tau}{\hbar}(\sigma_{x})_{w}<<1.
\end{equation}

\section{Beyond the standard weak measurement regime}

\begin{figure}[tb]
{\rotatebox{270}{\resizebox{8.0cm}{10.0cm}{\includegraphics{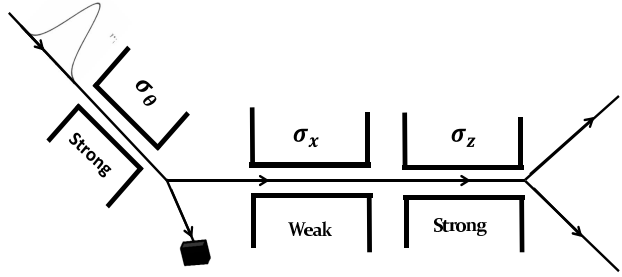}}}}%
\caption{A series of Stern-Gerlach setup for implementing weak measurements of the spin
operator $\hat{\sigma}_{x}$. The three SG setups account respectively for the state
preparation, the weak measurement, and the post selection .}
\end{figure}

\subsection{General remarks}

The usual weak measurement scheme that was presented in Sec.\ II involves
rough approximations, some of which can usually be controlled (coupling value,
width of device states), others, like the loss of unitarity due to the
asymptotic expansion have consequences that are more delicate to tackle with in
definite situations. Moreover there remains a whole continuous range of
regimes for which the approximations made in the standard WM derivation do not
hold, though the situation is still far from the ideal strong measurement
limit. These intermediate non-ideal regimes that can be referred as
`semi-weak' situations are considered in the present Section.

We first give the exact solution for a non-ideal Stern-Gerlach, that is
solving the coupled Schrodinger equations for a wave packet of a spin 1/2
particle without making any specific assumptions for the coupling strength and
the device states.\ We will then see how semiweak outcomes are obtained when a
strong measurement is added after the neutron has emitted the non-ideal
Stern-Gerlach. A crucial quantity that we introduce is the overlap $I$ of the
wave packets (recall the wave packets play here the r\^{o}le of the device
states) that quantifies the strength of a given measurement situation.\ We
will indeed see that $I\rightarrow0$ yields the strong ideal measurement
regime, while $I\rightarrow1$ corresponds to the usual AAV weak measurement
scheme. For an arbitrary value of $I$ the superposition of the device states
results in a a distribution yielding the semiweak values. The different types
of behavior of the pointer that can be obtained (one or several semiweak
values, eccentric semiweak values or semiweak values falling between the
eigenvalues, exact weak values that take into account details of the
interaction Hamiltonian) will also be discussed.\ The illustrations are deferred to Sec.\ IV.

\subsection{Neutrons passing through a non-ideal SG setup: exact solutions}

The setup is the same as the one discussed in Sec.\ II (see Fig.\ 1). We omit
the state preparation procedure and consider the initial spin state as given.
Let a beam of neutrons, passing through the second SG setup be represented by
the total wave function
\begin{equation}
\Psi\left(  \mathbf{x},t=0\right)  \equiv\psi_{0}\left(  \mathbf{x}\right)
\left\vert \uparrow\right\rangle _{\theta}%
\end{equation}
where $\left\vert \uparrow\right\rangle _{\theta}=\cos\frac{\theta}%
{2}|\uparrow_{z}\rangle+\sin\frac{\theta}{2}|\downarrow_{z}\rangle$ is the
initial state of the system (i,e., the spin). The spatial wave function
$\psi_{0}(\mathbf{x})$ corresponds to a Gaussian wave packet which is
initially peaked at $\mathbf{x}=0$ at $t=0$, given by
\begin{equation}
\psi_{0}\left(  \mathbf{x}\right)  =\frac{1}{{(2\pi\delta^{2})}^{{3}/{4}}}%
\exp\left(  -\frac{\mathbf{x}^{2}}{4\delta^{2}}+i\frac{p_{y}y}{\hbar}\right)
\label{initpack}%
\end{equation}
where $\delta$ is the initial width of the wave packet. The wave packet moves
along the $+y$ axis with the initial momentum $p_{y}$ (see Fig.\ 1). The
inhomogeneous magnetic field$^{1}$ $\mathbf{B}=(bx,0,0)$
\footnote{This form of magnetic field is unphysical as it does not satisfy the Maxwell
equation $\nabla.\mathbf{B}=0$. We need at least another component to make it
divergence free \cite{home07}.  However on average the effect of
these  additional  field  components  can  be  neglected  under  proper
circumstances,  resulting  in  this  effective  field  usually  found  in
textbooks and also employed in all the previous works dealing with weak measurements
of spin.} is directed along the $x$-axis and confined between
$y=0$ and $y=d$. The interaction Hamiltonian is $H_{i}=\mu\widehat{\sigma
}.\mathbf{B}$ where as above $\mu$ is the magnetic moment of the neutron. As
the wave packet propagates through the SG magnet, in addition to the $+y$ axis
motion, the particles gain momentum along $\pm\ x$-axis due to the interaction
of their spins with the field. The time evolved total wave function at $\tau$
(transit time of the peak of the wave packet within the SG magnetic field
region) after the interaction of spins with the SG magnetic field is given by
\begin{align}
\Psi\left(  \mathbf{x},\tau\right)   & =\exp\left(  {-\frac{iH_{i}\tau}{\hbar
}}\right)  \Psi(\mathbf{x},t=0)\nonumber\\
& =\alpha\psi_{+x}(\mathbf{x},\tau)\otimes\left\vert \uparrow\right\rangle
_{x}+\beta\psi_{-x}(\mathbf{x},\tau)\otimes\left\vert \downarrow\right\rangle
_{x}\label{tevolved}%
\end{align}
where the device states $\psi_{+x}\left(  \mathbf{x},\tau\right)  $ and
$\psi_{-x}\left(  \mathbf{x},\tau\right)  $ are the two components of the
spinor $\psi=\left(
\begin{array}
[c]{c}%
\begin{array}
[c]{c}%
\psi_{+}\\
\psi_{-}%
\end{array}
\end{array}
\right)  $ which satisfies the Pauli equation and $\alpha=\frac{1}{\sqrt{2}%
}\left(  \cos\frac{\theta}{2}+\sin\frac{\theta}{2}\right)  $ and $\beta
=\frac{1}{\sqrt{2}}\left(  \cos\frac{\theta}{2}-\sin\frac{\theta}{2}\right)
$. Note that Eq.(\ref{tevolved}) is an entangled state between position and
spin degrees of freedom. The reduced density matrix of the system in the
$x$-basis representation can be written as
\begin{equation}
\rho_{s}=\left(
\begin{array}
[c]{cc}%
\alpha^{2} & \alpha\beta I\\
\alpha\beta I^{\ast} & \beta^{2}%
\end{array}
\right) \label{rdm}%
\end{equation}
where $I$ is the overlap%
\begin{equation}
I=\int_{v}\psi_{+x}^{\ast}\left(  \mathbf{x},\tau\right)  \psi_{-x}\left(
\mathbf{x},\tau\right)  d^{3}x.\label{idef}%
\end{equation}
that quantifies the weakness of the measurement. The inner product $I$
is in general complex but here in our case $I$ is always real and positive.
The value of the $I$ can range from $0$ to $1$ depending upon the choices of
the relevant parameters, such as, the degree of magnetic field ($b$), the width of the
initial wave packet ($\delta$) and the transit time through the field region
within SG setup ($\tau$). We calculate the analytical expressions of
$\psi_{+x}\left(  \mathbf{x},\tau\right)  $ and $\psi_{-x}\left(
\mathbf{x},\tau\right)  $ by solving relevant Schroedinger equations.

The two-component Pauli equation for $\psi_{+x}$ and $\psi_{-x}$ can then be
written
\begin{align}
i\hbar\frac{\partial\psi_{+x}}{\partial t} &  =-\frac{\hbar^{2}}{2m}\nabla
^{2}{\psi}_{+x}+\mu bx\psi_{+x}\label{decoupled}\\
i\hbar\frac{\partial\psi_{-x}}{\partial t} &  =-\frac{\hbar^{2}}{2m}\nabla
^{2}{\psi}_{-x}-\mu bx\psi_{-x}%
\end{align}
The solutions of the above two equations at $t=\tau$ upon exiting the SG are
as follows (for a detailed derivation, see Ref.\cite{home07})
\begin{align}
\psi_{+x}\left(  \mathbf{x};\tau\right)   &  =\frac{1}{\left(  2\pi\delta^{2}\right)  ^{\frac{3}{4}}}\exp\left[  -\frac{z^{2}+(y-\frac{p_{y}\tau
}{m})^{2}+(x-\frac{p_{x}^{\prime}\tau}{2m})^{2}}{4\delta^{2}}\right]
\nonumber\label{sol1}\\
&  \times\exp\left[  i\left\{  -\Delta+\left(  y-\frac{p_{y}\tau}%
{2m}\right)  \frac{p_{y}}{\hbar}+\frac{p_{x}^{\prime}x}{\hbar}\right\}
\right]
\end{align}%
\begin{align}
\psi_{-x}\left(  \mathbf{x};\tau\right)   &  =\frac{1}{\left(  2\pi\delta
^{2}\right)  ^{\frac{3}{4}}}\exp\left[  -\frac{z^{2}+(y-\frac{p_{y}\tau
}{m})^{2}+(x+\frac{p_{x}^{\prime}\tau}{2m})^{2}}{4\delta^{2}}\right]
\nonumber\label{sol2}\\
&  \times\exp\left[  i\left\{  -\Delta+\left(  y-\frac{p_{y}\tau}%
{2m}\right)  \frac{p_{y}}{\hbar}-\frac{p_{x}^{\prime}x}{\hbar}\right\}
\right]
\end{align}
where $\Delta=\frac{{p_{x}^{\prime}}^{2}\tau}{6 m \hbar}%
$,~~$p_{x}^{\prime}=\mu b\tau$, and the spreading of the wave packet is
neglected throughout the evolution.

Here $\psi_{+x}\left(  \mathbf{x},\tau\right)  $ and $\psi_{-x}\left(
\mathbf{x},\tau\right)  $ representing the spatial wave functions at $\tau$
correspond to the spin states $\left\vert \uparrow\right\rangle _{x}$ and
$\left\vert \downarrow\right\rangle _{x}$ respectively, with the average
momenta $\langle\widehat{p}\rangle_{\uparrow}$ and $\langle\widehat{p}%
\rangle_{\downarrow}$, where $\langle\widehat{p}\rangle_{\uparrow\downarrow
}=(\pm p_{x}^{\prime},p_{y},0)$. Within the magnetic field the neutrons gain
the same magnitude of momentum $p^{\prime}_{x}=\mu b\tau$ but the directions
are such that the particles with eigenstates $|\uparrow\rangle_{x}$ and
$|\downarrow\rangle_{x}$ get the drift along $+x$-axis and $-x$-axis
respectively, while the $y$-axis momenta remain unchanged.

From these analytical expressions of $\psi_{+x}\left(  \mathbf{x},\tau\right)
$ and $\psi_{-x}\left(  \mathbf{x},\tau\right)  $ given by Eqs.(\ref{sol1})
and (\ref{sol2}) it is straightforward to compute the inner product $I$ [Eq.
(\ref{idef})] given by
\begin{equation}
I=\exp\left(  -\frac{\mu^{2}b^{2}\tau^{4}}{8m^{2}\delta^{2}}-\frac{2\mu
^{2}b^{2}\tau^{2}\delta^{2}}{\hbar^{2}}\right) \label{inp}%
\end{equation}
which explicitly depends upon the choices of the parameters $b$, $\delta$ and
$\tau$.

Now, after emerging from this non-ideal SG magnet the neutrons represented by
the entangled state given by Eq.(\ref{tevolved}) enter to another SG setup
where a strong measurement to be performed and the neutrons are post-selected
in a specific spin state.

\subsection{Subsequent strong measurement: post-selection and final pointer
state}

For this purpose, we consider immediately after the wavepacket exits
the WM\ Stern-Gerlach a subsequent strong measurement of the spin observable
$\widehat{\sigma}_{z}$. The knowledge of the exact solutions allow to treat
the strong and weak measurements on the same footing. The approximate magnetic field in this case is
$\mathbf{B}=(0,0,b^{\prime}z)$ and $T$ is the time during which the
interaction occurs. A strong measurement requires the magnetic field $b^{\prime}$
to be sufficiently strong so that the spatial wavepackets emerging from this SG
setup are orthogonal, i.e., the relevant inner product, analog to Eq. (\ref{inp}) vanishes.
If the neutrons having the position-spin entangled state
given by Eq.(\ref{tevolved}) enters to final SG setup then the time evolved
state that exits from the setup can be written as
\begin{align}
\Psi(\mathbf{x},\tau+T) &  \mathbf{=}\ \frac{\alpha}{\sqrt{2}}\psi_{+x+z}(\mathbf{x}%
,\tau+T)\left\vert \uparrow\right\rangle _{z}+\frac{\alpha}{\sqrt{2}}\psi_{+x-z}(\mathbf{x}%
,\tau+T)\left\vert \downarrow\right\rangle _{z}\label{ttevolved}\\
&  +\frac{\beta}{\sqrt{2}}\psi_{-x+z}(\mathbf{x},\tau+T)\left\vert \uparrow\right\rangle
_{z}+\frac{\beta}{\sqrt{2}}\psi_{-x-z}(\mathbf{x},\tau+T)\left\vert \downarrow\right\rangle
_{z}\nonumber
\end{align}

The inner products $\left\langle \psi _{+x+z}\right\vert \left. \psi _{+x-z}\right\rangle $
and $\left\langle \psi _{-x+z}\right\vert \left. \psi _{-x-z}\right\rangle $ at $t=\tau +T$ vanish for strong measurements.
The states $\psi_{\pm x\pm z}(\mathbf{x},\tau +T)$ can be calculated by using
$\psi_{+x}(\mathbf{x},\tau)$ and $\psi_{-x}(\mathbf{x},\tau)$ given by
Eqs. (\ref{sol1}) and (\ref{sol2}) respectively as the initial position wave
functions. Note here that since the inhomogeneous magnetic field is only along
the $z$-axis, within this SG setup, the neutrons move freely along the two other
directions. Hence the analytical forms of the states $\psi_{\pm x\pm
z}(\mathbf{x},\tau +T)$, are given by
\begin{align}
\psi_{+x\ \pm z}\left(  \mathbf{x};\tau+T\right)   &  =\frac{1}{\left(
2\pi\delta^{2}\right)  ^{\frac{3}{4}}}\exp\left[  -\frac{\left(  z\mp
\frac{p_{z}T}{m}\right)  ^{2}+\left(  y-\frac{p_{y}(\tau+T)}{2m}\right)
^{2}+\left(  x-\frac{p_{x}^{\prime}\tau}{2m}-\frac{p_{x}^{\prime}T}{m}\right)
^{2}}{4\delta^{2}}\right] \nonumber\label{sol3}\\
&  \times\exp\left[  i\left\{  -\Delta-\Delta^{\prime}\pm\frac
{p_{z}^{\prime}z}{\hbar}+\left(  y-\frac{p_{y}(\tau+T)}{2m}\right)
\frac{p_{y}}{\hbar}+\frac{p_{x}^{\prime}}{\hbar}\left(  x-\frac{p_{x}^{\prime
}T}{2m}\right)  \right\}  \right]
\end{align}%
\begin{align}
\psi_{-x\ \pm z}\left(  \mathbf{x};\tau+T\right)   &  =\frac{1}{\left(
2\pi\delta^{2}\right)  ^{\frac{3}{4}}}\exp\left[  -\frac{\left(  z\mp
\frac{p_{z}T}{m}\right)  ^{2}+\left(  y-\frac{p_{y}(\tau+T)}{2m}\right)
^{2}+\left(  x+\frac{p_{x}^{\prime}\tau}{2m}+\frac{p_{x}^{\prime}T}{m}\right)
^{2}}{4\delta^{2}}\right] \nonumber\label{sol4}\\
&  \times\exp\left[  i\left\{  -\Delta-\Delta^{\prime}\pm\frac
{p_{z}^{\prime}z}{\hbar}+\left(  y-\frac{p_{y}(\tau+T)}{2m}\right)
\frac{p_{y}}{\hbar}-\frac{p_{x}^{\prime}}{\hbar}\left(  x+\frac{p_{x}^{\prime
}T}{2m}\right)  \right\}  \right]
\end{align}
where $\Delta^{\prime}=\frac{{p_{z}^{\prime}}^{2}T}{6 m \hbar}$
and $p_{z}^{\prime}=\mu b^{\prime}T$

Assume we post-select the neutrons having spin state $|\uparrow_{z}\rangle$
then the post-selected position space wave function is written using
Eq.(\ref{ttevolved}); given by
\begin{equation}
\Psi_{Post}(\mathbf{x},\tau+T)\mathbf{=}\frac{1}{\sqrt{2}}\left[ \alpha\psi_{+x+z}(\mathbf{x}%
,\tau+T)+\beta\psi_{-x+z}(\mathbf{x},\tau+T)\right]\label{xyzpost}%
\end{equation}
The states $\psi_{+x+z}(\mathbf{x},T+\tau)$ and $\psi_{-x+z}(\mathbf{x}%
,T+\tau)$ are separable in $x,y,z$ and can thus be written with obvious
notation as $\psi_{\pm x+z}(\mathbf{x},T+\tau)\equiv\psi_{\pm x}^{\prime}%
({x},T+\tau)\psi_{y}(y,T+\tau)\psi_{+z}(z,T+\tau)$. Eq.(\ref{xyzpost})
becomes
\begin{equation}
\Psi_{Post}(\mathbf{x},\tau+T)\mathbf{=}\frac{1}{\sqrt{2}}\ \psi_{y}(y,T+\tau)\psi
_{+z}(z,T+\tau)\left[  \alpha\psi_{+x}^{\prime}({x},T+\tau)+\beta\psi
_{-x}^{\prime}({x},T+\tau)\right] \label{xyzpost2}%
\end{equation}
We see the $y$ and $z$ dependent parts of the total wave functions do not play
any significant role in this context are thus integrated out. The motion
along the $\hat{x}-$axis within the final SG setup is free so
we can neglect the $T$ dependence of $\psi_{\pm x}^{\prime}({x},T+\tau)$. Thus
the final post-selected pointer wave function, depending only
on $x$ can be written as
\begin{equation}
\Psi_{post}^{\prime}({x},\tau)=\frac{1}{\sqrt{2}}\left[\alpha\psi_{+x}^{\prime}({x},\tau)+\beta
\psi_{-x}^{\prime}({x},\tau)\right]\label{xpost}%
\end{equation}
where
\begin{equation}
\psi_{\pm x}^{\prime}({x},\tau)=\frac{1}{\left(  2\pi\delta^{2}\right)
^{\frac{1}{4}}}\exp\left[  -\frac{(x\mp\frac{p_{x}^{\prime}\tau}{m})^{2}%
}{4\delta^{2}}\pm i\frac{p_{x}^{\prime}x}{\hbar}\right] \label{xpsi}%
\end{equation}
obtained from Eqs.(\ref{sol3}) and (\ref{sol4}).

The corresponding momentum space wave function is%
\begin{equation}
\Phi_{post}({p_{x}},\tau)=\frac{1}{\sqrt{2}}\left[\alpha\phi_{+x}({p_{x}},\tau)+\beta\phi_{-x}({p_{x}%
},\tau)\right]\label{ppost}%
\end{equation}
where $\phi_{\pm x}$ are obtained by taking the Fourier transform of
Eqs.(\ref{xpsi}), yielding, with $\phi_{0}$ given by Eq. (\ref{a2})%
\begin{equation}
\label{psol}
\phi_{\pm}(p_{x},\tau)=\phi_{0}({p_{x}\mp p_{x}^{\prime}})\exp\left(
-\frac{ip_{x}^{2}\tau}{2m\hbar}\pm\frac{ip_{x}(p_{x}^{\prime})^{2}}{2m\hbar
}-\frac{i{\mu}^{2}b^{2}\tau^{3}}{6m\hbar}\right)
\end{equation}
The momentum space solution (\ref{ppost}) is general in that no restriction on
the coupling constant or on the width of the probe states have been
introduced. We can now look at the device's momentum distribution in different
regimes that will be characterized by the value of $I$.

\subsection{Strong measurement limit: $I\approx0$}

Let us now assume we tune the relevant parameters ($b$, $\delta$ or $\tau$) so
that $I\rightarrow0$. In this case the reduced density matrix given by Eq.
(\ref{rdm}) of the system becomes
\begin{equation}
\rho_{s}=\left(
\begin{array}
[c]{cc}%
\alpha^{2} & 0\\
0 & \beta^{2}%
\end{array}
\right)  .
\end{equation}
which is diagonal implying the strong measurement limit in that there is a
one-to-one correspondence between the pointer state $\phi_{+x}^{\prime}$
(resp. $\phi_{-x}^{\prime}$) and the spin state $\left\vert \uparrow
\right\rangle _{x}$ (resp. $\left\vert \downarrow\right\rangle _{x}$). Hence,
in this case we expect two spatially separated peaks pointing at the
eigenvalues of the spin observables $\widehat{\sigma}_{x}$. In Fig.(2a) we
depicted the device momentum distribution given by Eq.(\ref{ppost}) where we
find the peaks at $\pm p_{x}^{\prime}$.

\subsection{Standard weak measurement limit: $I\approx1$}

The WM limit is just the opposite limiting situation of strong measurement.
From Eq.(\ref{inp}) we can see for example that for a fixed $\delta$ we can
choose the other parameters $b$ and $\tau$ such that $I\approx1$ is obtained.
The reduced density matrix of the system becomes
\[
\rho_{s}\approx\frac{1}{2}\left(
\begin{array}
[c]{cc}%
\alpha^{2} & \alpha\beta\\
\alpha\beta & \beta^{2}%
\end{array}
\right)
\]
i.e., the coherence in the system is considered to be mostly unaffected. There
is no correspondence between the system and the device states. This leads to
the usual weak measurement scheme proposed by AAV. Indeed let us start with
the pointer position space wave function given by Eq.(\ref{xpsi}) since the
coupling is between the spin and the position variable. If the parameters $b$
and $\tau$ are sufficiently small the value of $I$ could be close to unity for
a fixed $\delta$. In this case we can neglect the higher order terms of
$b\tau$ keeping only first order term. We can then write the states $\psi_{\pm
x}^{\prime}\left(  \mathbf{x};\tau\right)  $ given by Eqs.(\ref{xpsi}) as
follows
\begin{equation}
\psi_{\pm x}^{\prime}\left(  {x};\tau\right)  \approx\psi_{0}(x)\left(  1\pm
ip_{x}^{\prime}x/\hbar\right) \label{appwfn}%
\end{equation}

In this limit Eq.(\ref{xpost}) becomes
\begin{equation}
\label{ppp}
\Psi_{post}^{\prime}({x},\tau)\approx \frac{\psi_{0}(x)}{\sqrt{2}}\left[  \alpha\left(
1+ip_{x}^{\prime}x/\hbar\right)  +\beta\left(  1-ip_{x}^{\prime}%
x/\hbar\right)  \right]
\end{equation}
Putting the values of $\alpha$, $\beta$ and $\psi_{0}(x)$, and simplifying we
get
\begin{equation}
\label{psii1}
\Psi_{post}^{\prime}({x},\tau)\approx\frac{\cos\frac{\theta}{2}}{\left(
2\pi\delta^{2}\right)  ^{1/4}}\exp\left(  -\frac{x^{2}}{4\delta^{2}}%
+i\frac{p_{x}^{\prime}x\tan\frac{\theta}{2}}{\hbar}\right)
\end{equation}
The Fourier transform of Eq.(\ref{psii1}) gives the pointer wave function in
momentum space yielding the distribution%
\begin{equation}
|\Phi_{f}(p_{x})|^{2}\approx\cos^{2}\frac{\theta}{2}\left(  \frac{2\delta^{2}%
}{\pi\hbar^{2}}\right)  ^{1/2}\exp\left[  -\frac{2\delta^{2}(p_{x}%
-p_{x}^{\prime}\tan\frac{\theta}{2})^{2}}{\hbar^{2}}\right]
\end{equation}
which exactly matches the Eq.(\ref{aavpdis}) of the standard WM formulation.

\subsection{Nonideal situations: semiweak and exact weak values}

When the overlap $I$ lies in the intermediate range $0<I<1$ an imperfect
coherence remains in the system state. After post-selecting on a given final
spin state, the pointer distribution is obtained from Eq. (\ref{xpost}),
therefore displaying an interference between the overlapping meter
wavefunctions correlated with orthogonal spin states.

The pointer behavior depends on the details of the interference: by monitoring
$I$, eccentric pointer shifts can be obtained in this intermediate range.\ We
coin these resulting shifts as `semi-weak' values. Depending on the resulting
interference, that can be tuned by changing the values of the parameters (such
as $b,$ $\delta$ or $\tau$) the pointer distribution can display a single
maximum, akin to the usual weak measurement regime with the additional feature
that this maximum does not necessarily appear at the AAV weak value given by
Eq. (\ref{aavwvalue}), but can be chosen by fine-tuning the parameters. Other
behaviors of the pointer can be obtained, in particular profiles with two
maxima, similar to the two peaks that characterize the strong limit, but with
the maxima shifted far from the eigenvalues.

Even in the usual $I\approx1$ weak measurement regime pointer shifts can be
obtained in cases in which the usual WM formalism does \emph{not} predict any,
such as in situations in which the pre- and post-selected states are
orthogonal. More importantly when the oscillating terms in Eqs. (\ref{xpsi})
play a role (that is when the wavelength of the oscillation becomes of the
order of magnitude of the distribution width), the meter distribution becomes
significantly different than what is predicted by the usual WM formalism. The
features produced by these `exact' weak values will be illustrated below.

It is interesting to note that several recent works \cite{wu,geszti,nakamura,zhu} have also
attempted to go beyond the usual WM formalism in the weak regime. Wu and Li \cite{wu} took
the asymptotic expansion to second order (one order beyond AAV's derivation) and
showed this was sufficient to treat the orthogonal pre and post-selected states case.
These results were extended in Ref. \cite{nakamura} where it was shown that the
asymptotic expansion can be resummed when weak values of projectors are considered.
A different but equivalent approach \cite{geszti,zhu} follows the early method of Sudarshan and collaborators \cite{duck}
consisting in formally expanding the preselected state in terms of the eigenstate of the weakly measured
operator. This allows to recover the standard WM in the limit of weak couplings. The main difference of all these
approaches with our present work lies not only in the method employed -- we explicitly
solve the full Schroedinger equation of the entire system (measured subsystem and pointer apparatus) --,
but also in the fact that by doing so, we take into account the motion of the system and of the pointer,
generated by the respective self-Hamiltonians.
The strong, weak or non-ideal nature of the measurement is then seen to depend on characteristics of the full solution,
not only on the coupling strength. For the weak measurement of spin in a Stern-Gerlach developed here,
the different parameters that determine the non-ideal character of the measurement were
encapsulated in the quantity $I$.

\section{Illustrations and Applications}

\subsection{Semiweak value distributions}

We give here an illustration of the properties of the meter state
distributions in the non-ideal regime giving rise to the semiweak values
examined in Sec. III. For definiteness the examples given will correspond to
the explicit formulas given above, involving the initial wavefunction
Eq.(\ref{inwfn}), with the initial spin in state $\left\vert \uparrow
\right\rangle _{\theta}$, an intermediate weak or semiweak measurement of the
system (spin) observable $\hat{\sigma}_{x}$, and a final post-selection along
$\left\vert \uparrow\right\rangle _{z}$.

\begin{figure}[t]
\includegraphics[height=7cm]{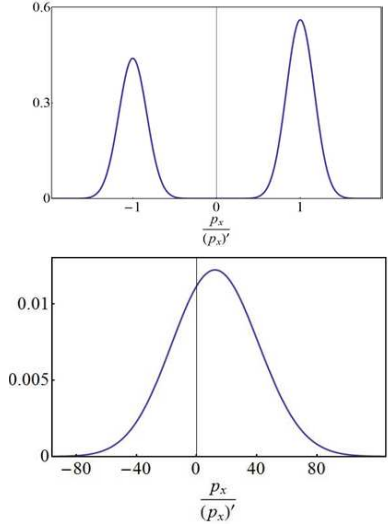}%
\caption{(Color online)The momentum distribution $\vert$$\Phi_{post}$(p$_{x}$,$\tau$)$\vert$$^{2}%
$(Eq.\ref{ppost}) is plotted when: (a) $I\approx0$ (ideal projective measurement
situation); and (b)$I\approx1$ (weak measurement situation). The values of
relevant parameters $b=100$G/cm, $\tau$=1.4$\times$10$^{-6}$s, and $\delta$ is 1cm
and 1/50cm for (a) and (b) respectively. The weak value is ($\sigma_{x}$%
)$_{w}$=16.2 for $\theta$=173.5$^{o}$.}%
\end{figure}

\subsubsection{Strong and weak limits}

Consider first the strong and weak limits. A typical example is given in Fig.
2, where the device state distributions are plotted in configuration
($x$) space as well as in momentum ($P$) space (recall the
only non-trivial axis is the one along $x$, the WM\ axis). For the strong case
the post-selection along $\left\vert \uparrow\right\rangle _{z}$ is applied
after a first strong measurement along $\hat{\sigma}_{x}$ that couples
perfectly each wavepacket (meter state) with the the relevant spin state
$\left\vert \uparrow\right\rangle _{x}$ or $\left\vert \downarrow\right\rangle
_{x}$. In $\mathbf{x}$ space this yields a single broad structure at the
eigenvalue of the $\hat{\sigma}_{z}$ measurement, whereas in $\mathbf{P}$
space two peaks are visible(see, Fig.\ 2(a)), each corresponding to the
direction of the momentum along the $x$-axis: the strong measurement condition
correlates perfectly this momentum with the spin state.

In the weak limit, a single peak is still visible in $x$ space, but
in $P$-space the two interfering Gaussians overlap almost perfectly
($I\approx1$).\ The result is itself (approximately \cite{duck}) a Gaussian,
as predicted by the WM formalism, but with a maximum displaced at the weak
value $p_{x}^{\prime}\tan\frac{\theta}{2}$ (see Fig.\ 2(b)).\ The meter has
therefore a broad distribution centered at a value several times higher than
the spin eigenvalue with the single peak
 pointing here at  $(\sigma_{x})_{w}=16.2$ for $\theta=173.5^{o}$..

\subsubsection{Semi-weak values}

We first illustrate in Fig.\ 3 the generic situation as $I$ is varied from the
ideal strong situation ($I\approx0$) to the weak limit ($I\approx1$). We see
that the meter distribution goes smoothly from the strong type of behavior to
the one that characterizes the weak limit.\ Note however that the maximum does
not appear anymore at the eigenvalues nor at the usual weak value Eq.
(\ref{aavwvalue}) but at the semiweak values, that is the maximal values of the
distribution obtained from Eq. (\ref{ppost}). In Fig. 3(c) for example
the left peak is shifted to $-14.2$ while the weak value is $(\sigma_{x}%
)_{w}=13.3$ (shown in Fig. 3(d)). This semi-weak value can be chosen at will by tuning the different
parameters; approximate analytical expressions may be obtained by expanding
$\left\vert \Phi_{post}({p_{x}},\tau)\right\vert ^{2}$ in terms of the
relevant small parameters ($\tau$, $b$ ,... or a combination thereof).

\begin{figure}[tb]
{\rotatebox{0}{\resizebox{10.0cm}{10.0cm}{\includegraphics{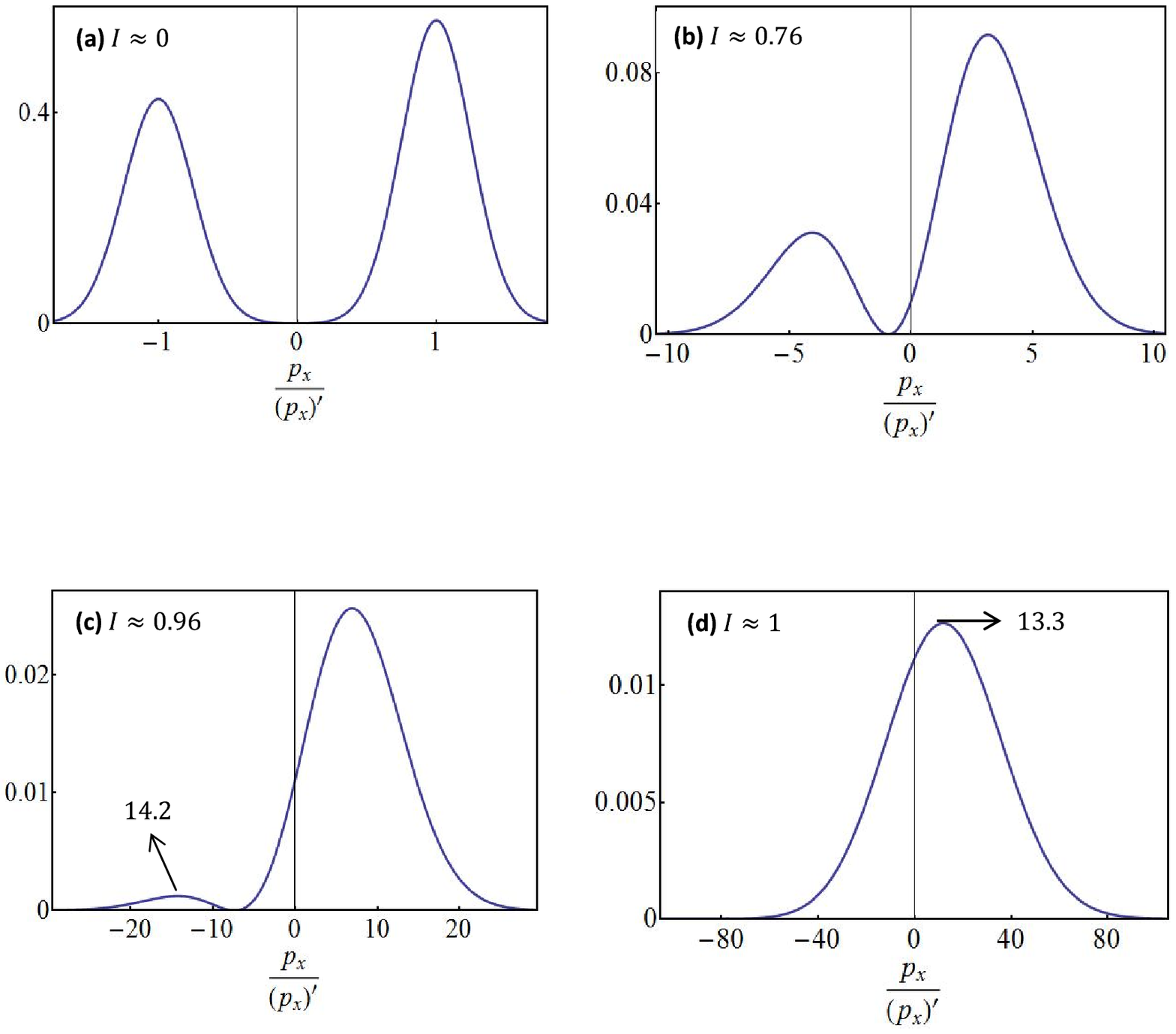}}}}%
\caption{(Color online)The momentum distribution $|\Phi
_{post}(p_{x},\tau)|^{2}$(Eq.\ref{ppost}) is plotted for four different values of
the inner product$(I)$ depending upon the suitable choices of the parameters.
Here (a) and (d) represents the strong and weak measurements, and (b)-(c)
correspond the semiweak measurement situations. The weak value is $(\sigma
_{x})_{w}=13.3$ for $\theta=171^{o}$.}%
\end{figure}

\subsubsection{Exact weak values}

A different feature of the exact formalism developed above concerns the
existence of a meter distribution peaked at eccentric values in the usual weak
regime (as far as the weakness of the coupling and the meter states are
concerned), but for which the usual AAV formalism does not yield any
result.\ For example when the initial and post-selected states are orthogonal
the usual weak value given by Eq.\ (\ref{aavwvalue}) is undefined. This
undefiniteness stems from partially expanding the exponential after the
post-selection [see Eq. (\ref{a6})]. An exact treatment does not have this
problem. Fig.\ 4 shows the exact meter momentum distribution for orthogonal
pre and post-selected states, for parameters near the $I\approx1$ limit. In
this situation, the post-selected meter state in momentum space is given by%
\begin{equation}
\label{ppostortho}
\Phi_{post}({p_{x}},\tau)=\frac{1}{{2}}\left[  \phi_{+x}({p_{x}}%
,\tau)-\phi_{-x}({p_{x}},\tau)\right]
\end{equation}
The analytical form of the momentum distribution near the $I\approx1$ limit is readily
obtained (up to a constant factor) from the Fourier transform of Eq.(\ref{ppp}) as
\begin{equation}
\label{orthoexact}
\left|\Phi_{post}({p_{x}},\tau)\right|^{2}\propto  p_{x}^{2}exp\left(-\frac{2 p_{x}^{2}\delta^{2}}{\hbar^{2}}\right)
\end{equation}
This distribution has two peaks at $p_{x}=\pm\hbar/\sqrt{2}\delta$ as illustrated in Fig.4.
The type of feature illustrated in Fig. 4 was first obtained theoretically in an approximate form
(by neglecting the evolution of the pointer inside the Stern-Gerlach) by Duck \emph{et al.}\cite{duck},
and more recently revisited in some of the works \cite{wu,geszti,nakamura,zhu} going beyond the standard WM (eg in
Ref. \cite{wu} a formula for orthogonal pre- and post-selected states is obtained
by considering the interaction up to second order).
Experimentally the feature was observed quite early \cite{ritchie}. In a more recent work\cite{starling2}, the electric
field analogue of Eq.(\ref{orthoexact}) was derived in the framework of a phase-amplification technique using classical wave optics.
Note that the exact weak value is
perfectly finite whereas from the definition given by Eq.(\ref{aavwvalue}) one
might be tempted to conclude (incorrectly as this would violate the conditions given by
Eq.(8) under which the AAV approximation holds) that the weak value can be
arbitrarily large as the pre and post-selected states become orthogonal.

\begin{figure}[tb]
{\rotatebox{0}{\resizebox{5.0cm}{4.0cm}{\includegraphics{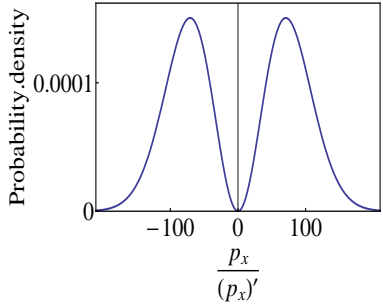}}}}%
\caption{{(Color online)\protect\footnotesize The momentum distribution $|\Phi_{post}%
(p_{x},\tau)|^{2}$(Eq.\ref{ppostortho}) is plotted when $I\approx1$ but the pre- and
post-selected states are orthogonal - a situation in which the standard weak value is
undefined.}}%
\end{figure}

\begin{figure}[tb]
{\rotatebox{0}{\resizebox{5.0cm}{4.0cm}{\includegraphics{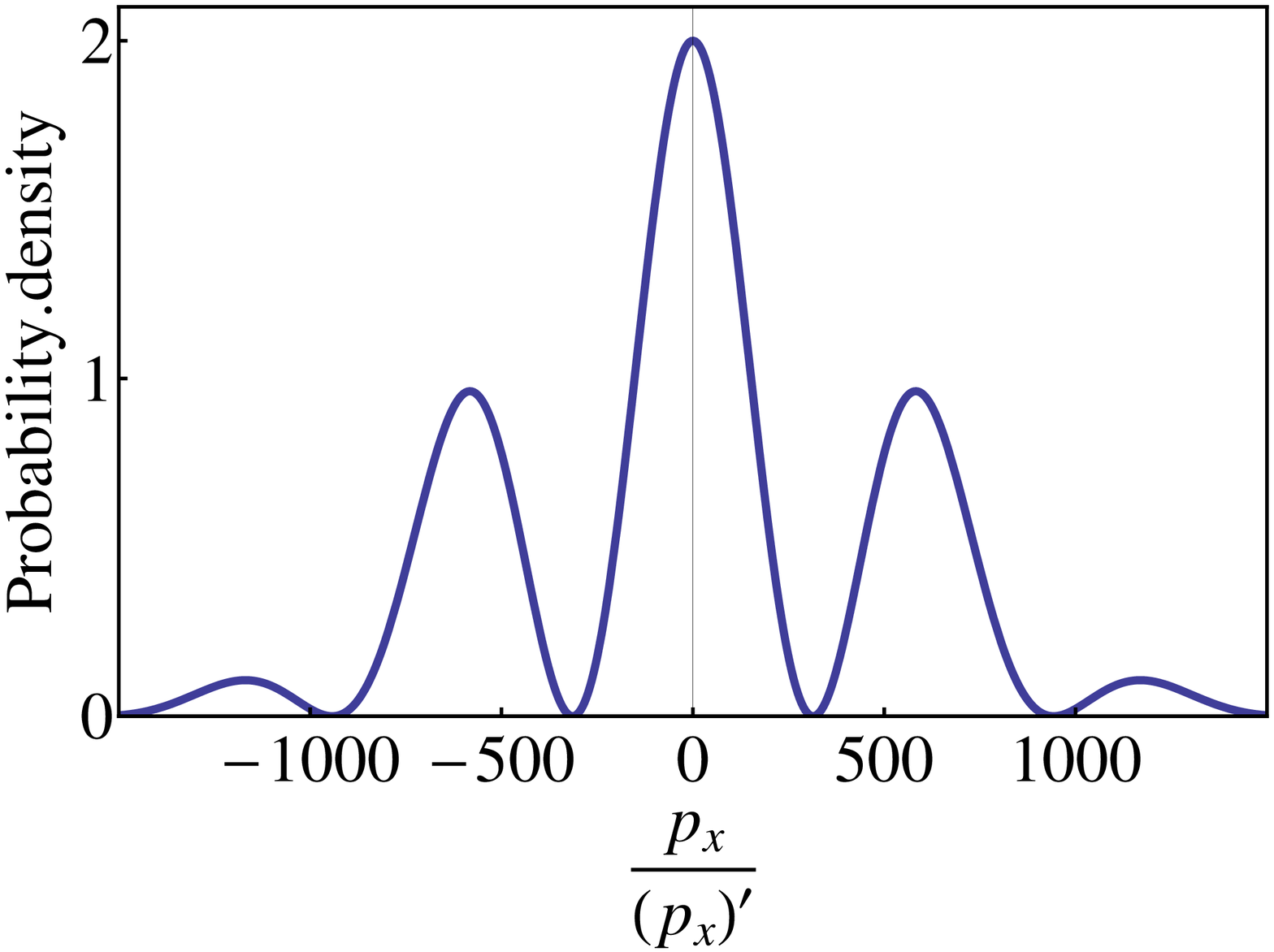}}}}%
\caption{{(Color online)\protect\footnotesize The momentum distribution $|\Phi_{post}%
(p_{x},\tau)|^{2}$ [Eq. (\ref{ppost})] is plotted when the  pre- and
post-selected states are identical. The values of
relevant parameters $b=0.001$G/cm, $\tau$=1.4$\times$10$^{-2}$s, and $\delta$ is $5\times 10^{-2}$ cm
.}}%
\end{figure}

\begin{figure}[tb]
{\rotatebox{0}{\resizebox{5.0cm}{4.0cm}{\includegraphics{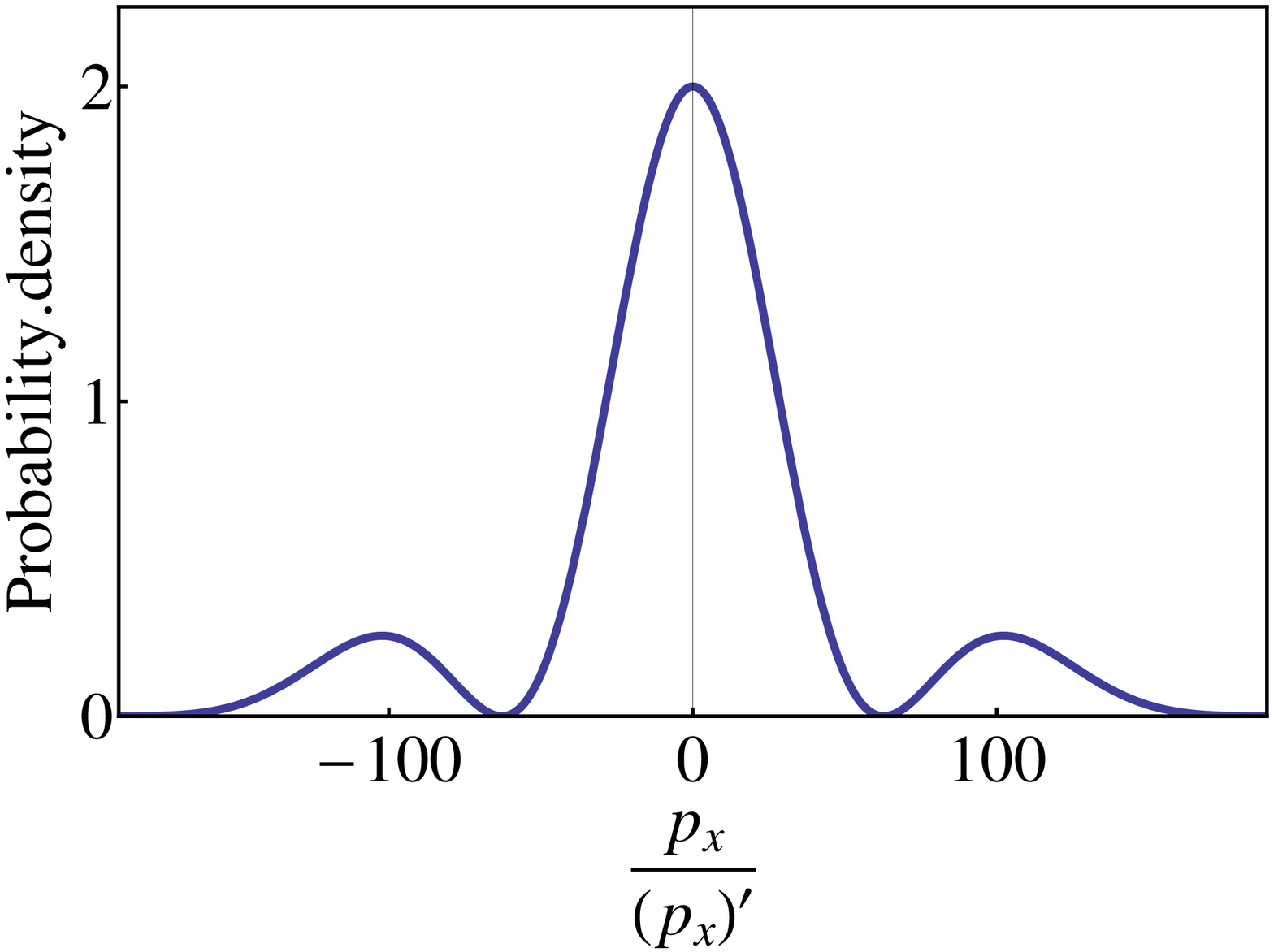}}}}%
\caption{{(Color online)\protect\footnotesize The momentum distribution $|\Phi_{post}%
(p_{x},\tau)|^{2}$ [Eq. (\ref{ppost})] is plotted when the pre- and
post-selected states are identical. The values of
relevant parameters $b=0.001$G/cm, $\tau$=1.4$\times$10$^{-2}$s, and $\delta$ is $10^{-3}$ cm
.}}%
\end{figure}

Another instance in which the 'exact' weak value differs from the one defined
in the asymptotic treatment appears when the oscillating terms of the exact
solutions (Eq.\ref{psol}) play a r\^{o}le.\ This happens when the
wavelength of these terms become comparable to the width of the momentum
distribution. A first example is shown in Fig.\ 5, when the pre- and
post-selected states are identical. Then according to its definition
(\ref{aavwvalue}) the weak value should simply be given by the average of the
weakly measured observable in the initial state. It can be seen in Fig.\ 5
that this rule is spoiled by the oscillating terms, that create a multiple
peak structure with eccentric maxima. By tuning the parameters entering the
exact wavefunction, the wavelength of the oscillating terms can be reduced,
yielding a momentum distribution with multiple peaks, as shown in Fig.\ 6.
 The origin of the multiple peaks is the interference between the
two meter states emerging from the second SG setup along two opposite directions
carrying the opposite phases. These phases (in momentum space) arise because of the separation of the two
wavepackets in configuration space due to the spin-inhomogeneous magnetic
field interaction occurring along opposite directions.
That shows the necessity of the solution of the Schroedinger equation
using the full Hamiltonian instead of taking only the interaction Hamiltonian or keeping only a
few terms of the asymptotic expansion of the interaction coupling.
Note here that the average momentum that has been the main interest
in several works\cite{wu,geszti,nakamura,zhu} is not necessarily relevant
when there are multiple peaks in the momentum distribution.

\subsection{Application: Distinguishing between ``identical" density matrices}

As an illustration of the practical usefulness of the exact scheme for
weak values reported in this work, we give here an application to a
quantum information task, namely the possibility of distinguishing between
"identical" density matrices. Consider the following situation: Alice prepares
neutral spin 1/2 particles in some state, either $\rho_{x}$ or $\rho_{z}$ and
sends them to Bob, whose goal is to guess the state. According to elementary
quantum mechanics the spin density matrices%
\begin{equation}
\rho_{x}\equiv\frac{1}{2}\left\vert \uparrow_{x}\right\rangle \left\langle
\uparrow_{x}\right\vert +\frac{1}{2}\left\vert \downarrow_{x}\right\rangle
\left\langle \downarrow_{x}\right\vert
\end{equation}
and%
\begin{equation}
\rho_{z}\equiv\frac{1}{2}\left\vert \uparrow_{z}\right\rangle \left\langle
\uparrow_{z}\right\vert +\frac{1}{2}\left\vert \downarrow_{z}\right\rangle
\left\langle \downarrow_{z}\right\vert
\end{equation}
are identical and thus undistinguishable. We must therefore give an additional
condition: we assume Alice sends successive spins of alternate signs, ie Alice
sends either $\xi=\{\left\vert \uparrow_{x}\right\rangle ,\left\vert
\downarrow_{x}\right\rangle ,\left\vert \uparrow_{x}\right\rangle ,\left\vert
\downarrow_{x}\right\rangle ...\}$ or $\zeta=\{\left\vert \uparrow
_{z}\right\rangle ,\left\vert \downarrow_{z}\right\rangle ,\left\vert
\uparrow_{z}\right\rangle ,\left\vert \downarrow_{z}\right\rangle ...\}$, each
of the sets $\xi$ and $\zeta$ giving rise to a specific realization of
$\rho_{x}$ or $\rho_{z}$ respectively. Bob must guess as fast as possible,
that is by processing the lowest number of particles, whether Alice is sending
$\xi$ or $\zeta$.

\begin{figure}[h]
\includegraphics[height=7.5cm]{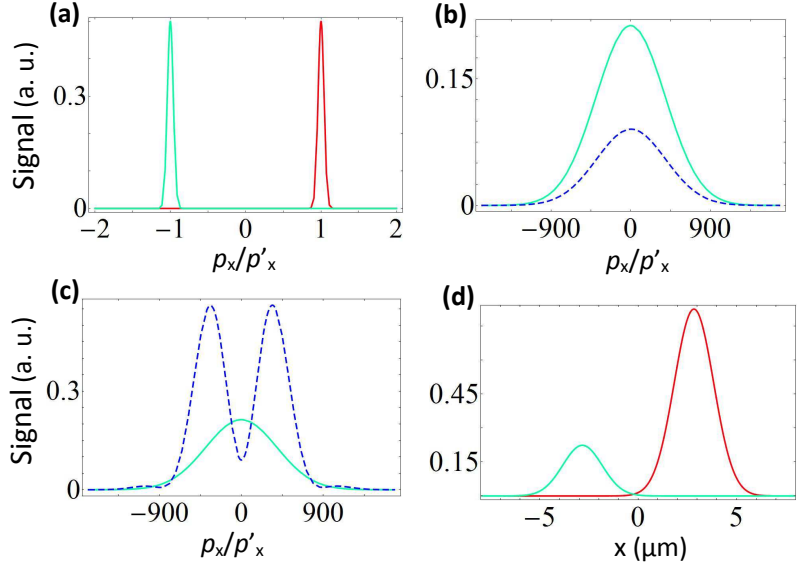}
\caption{(Color online)Meter distributions (in arbitrary units) for the task described in Sec. IV.2
for strong [(a)], standard weak [(b)] and exact weak [(c)-(d)] measurements, with the
magnetic field strength fixed in all cases at $b=0.02$ G/cm and $\tau=0.07$ s.
In (a)  $\widehat{\sigma }_{x}$ is measured strongly ($\delta=1$ cm) and the two peaks
appear at the corresponding eigenvalues, irrespective of whether $\xi$ or
$\zeta$ is being sent. (b) shows the meter distribution corresponding to $\xi$ (dashed, online blue curve)
and to $\zeta$ (solid, online green curve) for a standard WM  of  $\widehat{\sigma }_{x}$ (with $\delta=10^{-4}$ cm)
 and a post-selection along  $55^{\circ}$. (c) shows the same situation displayed in (b)
but with the exact WM solutions. (d) shows the \emph{configuration} space
meter distributions corresponding to the exact WM momentum space distributions shown in (c) with the same post-selection
angle. When Alice launches $\xi$ the left (online green) and right (online red) peaks are obtained for
$\left\vert\downarrow_{x}\right\rangle$ and $\left\vert \uparrow_{x}\right\rangle$
 respectively, whereas for $\zeta$, $\left\vert\downarrow_{z}\right\rangle$ and $\left\vert \uparrow_{z}\right\rangle$
 can both hit the left and right peaks.}
\label{applidm}
\end{figure}

With strong measurements, the best Bob can do is to measure either
$\widehat{\sigma}_{x}$ or $\widehat{\sigma}_{z}$ and examine whether two
successive measurements have identical signs. Suppose for instance Bob chooses
to measure $\widehat{\sigma}_{x}$; the pointer displays two sharp peaks, one
positive and one negative (see Fig. \ref{applidm} (a)). Then if $\xi$ is sent
the sign of successive outcomes will alternate, whereas if $\zeta$ is sent
each outcome is equiprobable.\ So Bob's strategy will be to observe whether
the consecutive measurements have alternating outcomes, in which case he will
conclude it is likely Alice is sending $\xi$. Indeed denoting by $k$ the
number of particles that have already been seen to be displaying an alternate
series, the probability of continuing with this series for the $(k+1)$th
particle if $\zeta$ has been sent becomes $2^{-k}$ and decrease rapidly with
$k$.

With standard weak measurements the problem for Bob is that he ignores the
preselected state: he must guess whether Alice is sending $\xi$ or $\zeta$ by
obtaining meter distributions that are very broad in momentum space,
irrespective of the weak measurement and post-selection Bob chooses. Typically
all the meter distributions have almost identical profiles, but different
heights, which is how $\xi$ and $\zeta$ can be distinguished (see Fig.
\ref{applidm} (b)). So despite the fact that the meter distributions will
indeed be different, in practice Bob will need a great number of particles in
order to discriminate $\xi$ from $\zeta$.

However by following the exact treatment in the non-ideal case, Bob can set
(by changing the SG magnetic field strength and passage time) the interference
between the meter states in momentum space so that the detection of $\xi$ and
$\zeta$ result in totally different probability distributions. Moreover by
weakening the coupling constant, Bob can still keep markedly different
profiles in \emph{momentum} space (see Fig. \ref{applidm} (c)) while obtaining
non-overlapping profiles for the meter in \emph{configuration} space.\ The
advantage is obvious: one has exactly the same information as the one obtained
with strong measurements (but in configuration space, see Fig. \ref{applidm}
(d)), and in addition as more particles are detected the detection curve in
momentum space unambiguously reveals whether $\xi$ or $\zeta$ is being sent.
It is crucial to note that the two distinctly positioned peaks visible in
configuration space are a feature of the exact solutions -- in the standard WM
formalism the configuration space wavefunctions are identical up to some
global factor. In practice \cite{pan matzkin prep}, each odd numbered result
is registered separately from the even numbered events, and this is done for
each of the two post-selected states.\ The procedure can be stopped when
enough events are registered so that the curves can be discriminated by
employing standard statistical tests minimizing the distance between the
expected curve and the obtained outcome \cite{pan matzkin prep}.

\section{Summary and Conclusions}
In this work we have derived the exact behavior of a measurement apparatus for arbitrary non-ideal measurements. This was done in the analytically tractable case of a spin measurement in Stern-Gerlach setups, for which the exact time-dependent wavefunctions can be obtained. This has allowed us to investigate the validity of the usual AAV weak measurement formalism, both in the intermediate regime, far from the usual WM limit, but also in the WM limit when the specific features of the system-meter interaction are explicitly accounted for. In doing so we have been led to introduce `semiweak' and `exact weak' values that describe the meter distribution and the associated eccentric values in these circumstances (including, but not limited to, the case of orthogonal pre- and post-selected states, for which the usual weak-value is undefined).

We have also seen through our exact approach that including the full Hamiltonian (rather than restricting the treatment to the interaction Hamiltonian) may have crucial consequences on the behavior of the eccentric values. In this sense, an exact approach brings additional features than those that can be retrieved by other recent works \cite{wu,geszti,nakamura,zhu} dealing with extending the original weak measurement approach: it is valid irrespective of the interaction strength and it yields fine details due to the specificities of the interaction. On the other hand, the practical applicability of an exact approach hinges on the ability to solve (analytically or numerically) the full Schr\"{o}dinger equation, a task that can become difficult for an arbitrary system. In all cases, going beyond the usual WM formalism as well as the relevance of performing exact calculations for a given experimental situation should be helpful in analyzing experimental results obtained in nonideal measurements.

\end{document}